\documentclass[sigconf]{acmart}
\AtBeginDocument{%
  }

\usepackage{algorithm}
\usepackage{algpseudocode}

\copyrightyear{2025}
\acmYear{2025}
\setcopyright{cc}
\setcctype{by}
\acmConference[KDD '25]{Proceedings of the 31st ACM SIGKDD Conference on Knowledge Discovery and Data Mining V.2}{August 3--7, 2025}{Toronto, ON, Canada}
\acmBooktitle{Proceedings of the 31st ACM SIGKDD Conference on Knowledge Discovery and Data Mining V.2 (KDD '25), August 3--7, 2025, Toronto, ON, Canada}
\acmDOI{10.1145/3711896.3737209}
\acmISBN{979-8-4007-1454-2/2025/08}

\begin{document}

\title{DV365: Extremely Long User History Modeling at Instagram}

\author{Wenhan Lyu}
\email{wenhanl@meta.com}
\orcid{0009-0002-9131-6491}
\affiliation{%
  \institution{Meta Platforms, Inc.}
  \city{Menlo Park}
  \country{USA}
}

\author{Devashish Tyagi\textsuperscript{†}}
\email{devashisht@openai.com}
\orcid{0009-0009-0757-2400}
\affiliation{%
  \institution{OpenAI}
  \city{San Francisco}
  \country{USA}
}

\author{Yihang Yang}
\email{yihangyang@meta.com}
\orcid{0000-0003-1984-5815}
\affiliation{%
  \institution{Meta Platforms, Inc.}
  \city{New York}
  \country{USA}
}

\author{Ziwei Li}
\email{ziweili@meta.com}
\orcid{0000-0002-1999-0011}
\affiliation{%
  \institution{Meta Platforms, Inc.}
  \city{Menlo Park}
  \country{USA}
}

\author{Ajay Somani\textsuperscript{†}}
\email{ajay250334@gmail.com}
\orcid{0009-0009-7350-7770}
\affiliation{%
  \institution{Independent}
  \city{New York}
  \country{USA}
}

\author{Karthikeyan Shanmugasundaram}
\email{karthikss@meta.com}
\orcid{0009-0003-0374-0071}
\affiliation{%
  \institution{Meta Platforms, Inc.}
  \city{New York}
  \country{USA}
}

\author{Nikola Andrejevic}
\email{hmnikola@meta.com}
\orcid{0009-0003-0576-1781}
\affiliation{%
  \institution{Meta Platforms, Inc.}
  \city{New York}
  \country{USA}
}

\author{Ferdi Adeputra}
\email{fadeputr@meta.com}
\orcid{0009-0001-1557-1004}
\affiliation{%
  \institution{Meta Platforms, Inc.}
  \city{New York}
  \country{USA}
}

\author{Curtis Zeng}
\email{curtiszeng@meta.com}
\orcid{0009-0008-5563-5229}
\affiliation{%
  \institution{Meta Platforms, Inc.}
  \city{New York}
  \country{USA}
}

\author{Arun K. Singh}
\email{arunksingh@meta.com}
\orcid{0009-0004-1991-0976}
\affiliation{%
  \institution{Meta Platforms, Inc.}
  \city{Menlo Park}
  \country{USA}
}

\author{Maxime Ransan}
\email{maxransan@meta.com}
\orcid{0009-0008-0396-1284}
\affiliation{%
  \institution{Meta Platforms, Inc.}
  \city{New York}
  \country{USA}
}

\author{Sagar Jain\textsuperscript{†}}
\email{jnsagar08@gmail.com}
\orcid{0009-0001-0296-3092}
\affiliation{%
  \institution{Independent}
  \city{Menlo Park}
  \country{USA}
}

\authornote{\textsuperscript{†}Work done while at Meta.}

\renewcommand{\shortauthors}{Lyu et al.}




\begin{abstract}
  Long user history is highly valuable signal for recommendation systems, but effectively incorporating it often comes with high cost in terms of data center power consumption and GPU. In this work, we chose offline embedding over end-to-end sequence length optimization methods to enable extremely long user sequence modeling as a cost-effective solution, and propose a new user embedding learning strategy, multi-slicing and summarization, that generates highly generalizable user representation of user's long-term stable interest. History length we encoded in this embedding is up to 70,000 and on average 40,000. This embedding, named as DV365, is proven highly incremental on top of advanced attentive user sequence models deployed in Instagram. Produced by a single upstream foundational model, it is launched in 15 different models across Instagram and Threads with significant impact, and has been production battle-proven for >1 year since our first launch.
\end{abstract}

\maketitle

\section{Introduction}

Large-scale recommendation systems are evolving fast driven by strong business needs. The most business critical recommendation models have been scaled for years and are often expensive large models. These models are resource-intensive during both training and deployment, including support systems such as feature extraction. Organizations are all resource-constrained, not only by capacity budgets but also by limited GPU supply in the market. This pushes us to scale models cautiously with ROI (return on investment) in mind.

Like previously found by Alibaba \cite{Pi_2019_MIMN}\cite{Ren_2019_HPMN} and Kuaishou \cite{chang2023twin}, we also found scaling user history length a compelling direction of improving recommendation quality. Currently in Instagram, recommendation models can access a user engagement history of up to 2k in length, bottlenecked by feature storage, extraction and processing costs. In our most advanced models, a user history of up to 500 in length is encoded by strong but costly attentive sequence encoder, HSTU \cite{zhai2024actionsspeaklouderwords}, which is bottlenecked by model training and serving GPU costs. 

Directly scaling up sequence lengths in existing models is not an option since we already reached our ROI limit after years of iterations, and attentive sequence models like HSTU have well-known challenges on sequence length scaling. We have 2 choices on the table: 1) End-to-end (E2E) model redesign with sequence length optimization methods \cite{cao2022samplingneedmodelinglongterm} \cite{chang2023twin} \cite{qi2020searchbasedusermodelinglifelong} \cite{Pi_2019_MIMN} \cite{Ren_2019_HPMN}; 2) Pre-train long user history into user embeddings and knowledge transfer to downstream models as features.

We decided to take the offline embedding approach for the following reasons:
\begin{enumerate}
    \item \textbf{Unlock Ambitious Length Scaling}: We target to unlock as much recommendation quality win as possible and we ended up scaling the combined sequence length to 70K in maximum and 40K on average.
    \item \textbf{Stable Interest Hypothesis}: We hypothesize that user interests consist of emerging interest and stable interest. The main incremental value of the long user history is in stable user interest, which does not change rapidly over time and needs a lower cadence of recurring training than production models with online training enabled.
    \item \textbf{Cross-domain Knowledge Sharing}: In Instagram, we have 20+ models serving different product surfaces like Reels, Feed, Explore, Stories, and across recommendation stages, including Retrieval, Early stage ranking (ESR), Late stage ranking (LSR), which is a common situation for many organizations. Doing the expensive long history computation only once and sharing with all models is a critical ROI attribute.
    \item \textbf{Feature Infra Costs Saving}: Real-time retrieving and processing extremely long sequences in every recommendation request, which is required by E2E methods, is a high cost and can be saved with offline embedding approach. 
\end{enumerate}

In this paper, we introduce an offline user embedding approach, with code name \textbf{DV365}, that uses a \textbf{Multi-slicing and Summarize (MSS)} strategy to encode user’s extremely long history into a user embedding. Considering the scale we are handling, we don’t apply attentive sequence modeling in the long sequence, but use multi-slicing to divide the long sequence into a set of sub-sequences and apply pooling. Multi-slicing and pooling produces 200 pooled embeddings. We also design a Funnel Summarization Arch as a user encoder to encode them into a condensed embedding. The user encoder module is embedded in a backbone simulation network similar to Instagram's production LSR model to simulate the E2E user encoder joint learning to help the user encoder produce a more consistent embedding for production models.

Our upstream model is trained and published in recurring manner to consume new training data, and store in a key-value store with 3 billion user to embedding pairs for downstream models.  

All offline embedding approaches has the following limitations, but we proved in this work that DV365 is a viable solution:
\begin{enumerate}
    \item \textbf{Freshness:} Model freshness is critical for recommendation quality \cite{Pinteest_Engineering_2022} to timely capture emerging data distribution. In Instagram, most critical models enabled online training \cite{Online_Machine_Learning_2024} to maximize freshness. Compared to E2E learning, offline embedding inevitably introduces delays between when the user history was encoded and when the signal is used in making recommendation decisions. To mitigate this, we design our embedding model objective to focus on “stable interest” which doesn’t change much over time and is robust to low freshness, and verified empirically that DV365 embedding quality has negligible regression as model delay increases.
    \item \textbf{Knowledge Transfer Efficiency and Generalizability:} Compared to user encoder trained E2E directly in production models, knowledge transferred from offline models inevitably has knowledge loss. Since our goal is to use a single foundational model to produce embedding for serving multiple downstream models, the knowledge loss casts a challenge on generalizability. We tackle this problem by training the embedding in a generic way with cross-surface data, and add adaptation modules in downstream models for domain adaptation. We prove the feasibility of this method using 15 downstream model adoptions with significant quality improvement.
\end{enumerate}

Our contributions can be summarized as follows. First, we introduce a multi-slicing and summarize user embedding training strategy that produces embeddings that give significant incremental values to production models with state-of-the-art (SOTA) user modeling components such as HSTU. We prove offline embedding as a high ROI solution for recommendation systems that want to unlock knowledge from extremely long user history.
Second, we have deployed DV365 embeddings to 15 production models in Instagram, Threads with a single embedding offering with significant impact.  This solution is time- and production-proven since our first launch in November 2023.

\section{Method}

\begin{figure*}[ht] 
    \centering \
    \includegraphics[width=\textwidth]{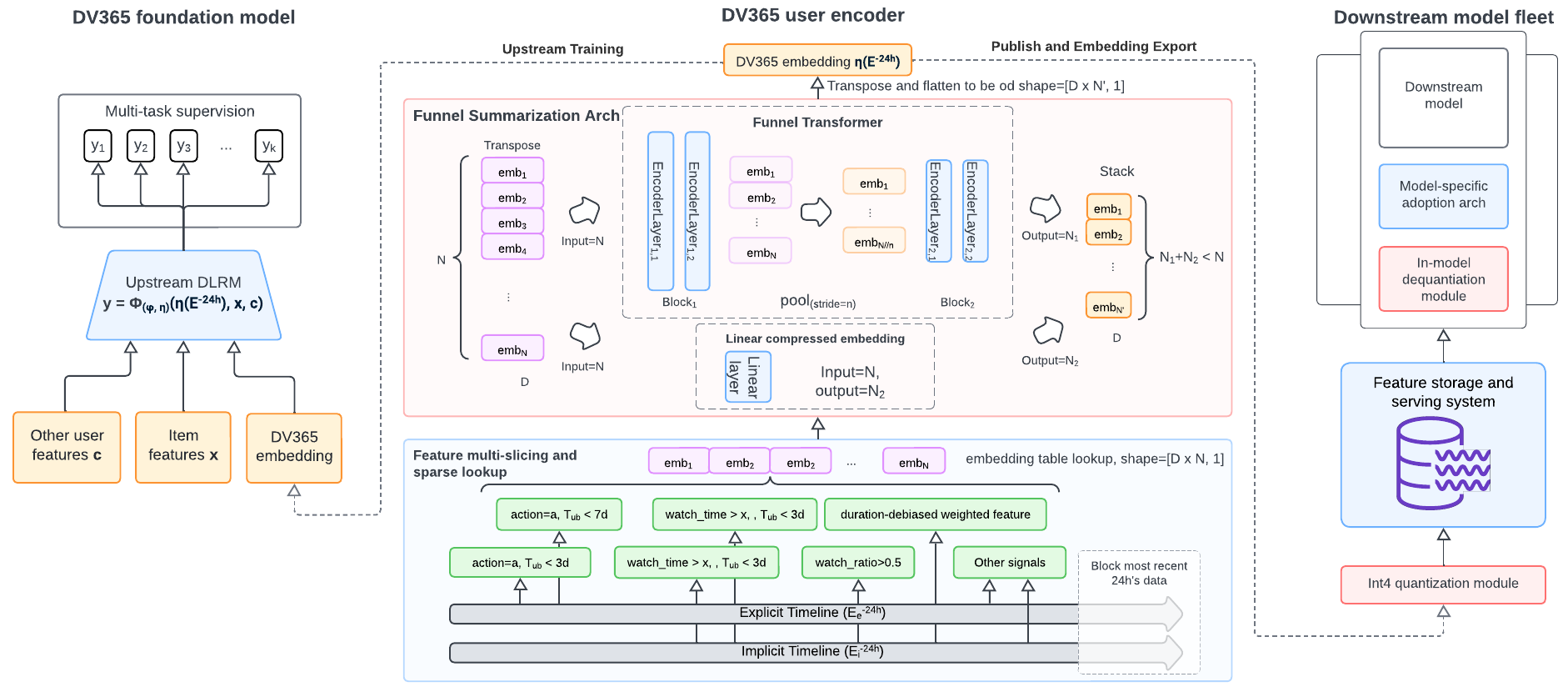} 
    \caption[Full model diagram]{
        Architecture of model training and embedding exporting.
}
    \Description[Full model diagram]{Full model diagram} 
    \label{fig:full_model_diagram} 
\end{figure*}

As shown in Figure \ref{fig:full_model_diagram}, the objective of our system is to build an offline foundational model that effectively pre-computes long user sequence information into a shareable user embedding to elevate the performance of a fleet of downstream models, agnostic of downstream model objectives and architectures.

\textbf{Step 1}: The upstream foundation model is trained and published recurringly, so that the user embedding is updated every few hours. The foundation model has 2 components: 
\begin{enumerate}
    \item \textbf{User encoder}: We use multi-slicing and summarization strategy in which we preprocess raw user interaction history into many sliced sub-sequences as categorical features, and summarize them into a compact user representation with a neural network. The output of this module becomes the user embedding used by downstream models.
    \item \textbf{Backbone Simulation Network}: To simulate the downstream use case and align user encoder training with recommendation model objectives, user encoder is embedded in a backbone simulation network with regular ranking objectives and other ranking features.
\end{enumerate}

\textbf{Step 2}: Downstream models incorporate the embedding with light-weight domain adaptation modules.

\subsection{Data Model and Preprocessing}

We design a unified user timeline (UniTi) which models user history as unified sequences of structured user interactions. Each user interaction contains fields of engaged media ID, author ID, event timestamp, action type (like, share, comment, etc.), video duration, user watch time, surface type, and media type.

The UniTi format can be described as follows. Let $\mathbf{m} = (e_1, e_2, e_3, ...)$ be a vector representing a media item, where element represents one media attribute such as the media id, author id, topic id, and media type. Based on user's action, we define two types of user timelines:
\begin{equation}
 \begin{split}
  &E_e = [..., (\mathbf{m}_i, \text{action\_type}_i, t_i), ...], i=0,1, ... \\
  &E_i = [..., (\mathbf{m}_j, \text{video\_duration}_j, \text{dwell\_time}_j, t_j), ...], j=0,1, ...
  \end{split}
  \label{eq:uniti_timelines}
\end{equation}
where $E_e$ is the \textbf{explicit timeline} (explicit user actions such as like, share), $E_i$ is the \textbf{implicit timeline} (user impression history of any user dwell time), and $i$ and $j$ are the respective index of the user engagements from a sampled data logger. 

Organizing user features as raw timelines ensures flexibility in feature engineering. Machine learning engineers can select any filtering criteria at the in-model preprocessing stage to create features on the fly. 

\subsection{Feature engineering}

We design UniTi features similarly to making rule-based sequence aggregation, in contrast to learnable aggregation used in sequence modeling methods like HSTU which is much more expensive.  

For the explicit timeline, we simply slice features by action types
\begin{equation}
  f_{\text{action}=a}(E_e) = \text{select}\ E_e\ \text{if}\ \text{action\_type} = a,
  \label{eq:explicit_features}
\end{equation}
so that a user's liked and commented media will have separate sequences. 
Similarly, user engagement in different logging sources (video or photo) will also be in different sub-sequences.

For the implicit timeline, we create sub-sequence features based on users' dwell time and the engaged videos' duration. We define 2 criteria based on dwell time and watch percentage: 
\begin{align}
  & f_{\text{watch\_time}}(E_i) = \text{select}\ E_i\ \text{if}\ \text{dwell\_time} \in [T_{\text{lb}}, T_{\text{ub}}];\label{eq:implicit_watch_time_features}\\
  & f_{\text{watch\_ratio}}(E_i) = \text{select}\ E_i\ \text{if}\  \frac{\text{dwell\_time}}{(\text{video\_duration})^\beta} \in [\alpha_{\text{lb}}, \alpha_{\text{ub}}], \label{eq:implicit_watch_ratio_features}
\end{align}
where $T_{\text{lb}}$, $T_{\text{ub}}$, $\alpha_{\text{lb}}$, $\alpha_{\text{ub}}$ are the lower and upper bounds of the watch time/ratio criteria, and $\beta$ is a tunable exponent. Having $\beta = 1$ corresponds to the normal watch percentage, and having $\beta < 1$ shifts the watch time distribution to be more fine-grained for shorter videos. 

To address the dwell time bias relative to video duration (users tend to watch longer for longer videos), we also created a video duration debiased weighted categorical feature: 
\begin{equation}
  \begin{split}
  & s(E_i) = \frac{\ln(\text{dwell\_time} + 1)}{\ln(\gamma\cdot\text{video\_duration} + 1)},\\
  & f_{\text{debiased\_wt\_score}}(E_i) = \text{select}\ (E_i, s(E_i))\ \text{if}\  s(E_i) > s_0,
  \end{split}
  \label{eq:debiased_wt_score_feature}
\end{equation}
where $s(\cdot)$ is the score formula, $\gamma$ and $s_0$ are tunable hyper-parameters. 

For the non-weighted categorical features, we also apply time-slicing:
\begin{equation}
  g_{T_{\text{ub}}}(f(E)) = \text{select}\ f(E)\ \text{if}\ \Delta t \in (0, T_{\text{ub}}]
  \label{eq:time_slicing},
\end{equation}
where $f(E)$ can be any non-weighted categorical feature generated from the explicit ($E_e$) or implicit ($E_i$) timelines, $\Delta t$ is equal to recommendation time minus the user's action time in the past, and $T_{\text{ub}}$ is the upper bound of the time bucket. In addition to full time horizon, we choose 2 additional time buckets $T_{\text{ub}}\in \{3\ \text{days}, 7\ \text{days}\}$, and apply time-slicing on all action-based explicit features (Eq. \ref{eq:explicit_features}) and the watchtime-based implicit features (Eq. \ref{eq:implicit_watch_time_features}).  

We performed exhaustive experiments and selected 200 derived features based on offline eval metrics improvements. 
The non-weighted categorical features are aggregated to be a single embedding per user using mean-pooling, and the weighted categorical features are aggregated using weighted mean-pooling (apply weighted sum divided by sum of all weights). 
We also enable embedding table sharing by grouping the categorical features by the following criteria: positive engagements (e.g, watch\_time > 15s) vs. negative engagements (e.g., watch\_time < 3s), implicit vs. explicit timelines, id type (media id, author id, topic id), non-weighted vs weighted, and product type. This setup forces each of the embedding tables to have its own learning focus, while maintaining a low cost in memory usage thanks to table sharing.

\subsection{Model architecture}

\subsubsection{Backbone Simulation Network}\label{sec:backbone_model}

The design goal of backbone simulation network is to create a training wrapper to supervise the user encoder to learn a module that can generate user embeddings that's effective in heterogeneous downstream models.

We use DLRM \cite{mudigere2022DLRM} as backbone network which is a multitask ranker $\mathbf{\Phi}$ performing the following task: given a user with organized UniTi timelines $E_i$ and $E_e$, targeting item features $\mathbf{x}$, and other user features $\mathbf{c}$, we aim to produce a score vector $\hat{\mathbf{y}} = \mathbf{\Phi}(E_i, E_e, \mathbf{x}, \mathbf{c})$ whose dimension is the same as the number of tasks, so that the total loss of all tasks is minimized. The loss is defined as cross entropy for binary tasks, and mean squared error (MSE) for regression tasks. The binary tasks loss is
\begin{equation}
  \begin{split}
  l_{\text{binary}}^{(k)} & = {\text{BCEloss}(\hat{y}^{(k)}, y^{(k)})}\\
  & = -\frac{1}{B}\sum_{j=1}^B \left[y_j^{(k)}\ln(\hat{y}_j^{(k)})+(1-y_j^{(k)})\ln(1-\hat{y}_j^{(k)})\right]
  \end{split}
  \label{eq:binary_loss}
\end{equation}
where $y_j^{(k)}$ is the ground truth label for binary task $k$ of the $j$th item in a mini batch, and $\hat{y}_j^{(k)}$ is the model's prediction. Our model has a single regression task predicting user's watch time, and its MSE loss is $l_{\text{reg}} = \frac{1}{B}\sum_{j=1}^B (z_j-\hat{z}_j)^2$. 
where $z$ and $\hat{z}$ is the ground truth and prediction of the user's dwell time. The total loss is 
\begin{equation}
  L_{\text{total}} =  \sum_k l_{\text{binary}}^{(k)} + a \cdot l_{\text{reg}},
  \label{eq:total_loss}
\end{equation}
where $a$ is a hyper-parameter to balance the scale of different loss types. 

\textbf{Distant Interest Prediction Objective} To ensure that the user embedding is stable, we force the UniTi user timeline to have a 24h gap before the interaction time, i.e., removing the most recent 24h information from the UniTi timelines $\mathbf{E}^{-24\mathrm{h}} = (E_i^{-24\mathrm{h}}$, $E_e^{-24\mathrm{h}})$. 
This 24h gap amounts to introducing a harder prediction task which enforces the model to learn a long-term user interest. As a result, the embedding is more robust against delays in training and data pipelines (Details in \ref{appendix:robustness_to_staleness}). The training objective then becomes
\begin{equation}
  \hat{\mathbf{\Phi}} = \mathrm{argmin}_{\mathbf{\Phi}}L_{\text{total}}(\mathbf{\Phi}(\mathbf{E}^{-24\mathrm{h}}, \mathbf{x}, \mathbf{c}), \mathbf{y}). 
  \label{eq:ranker_objective}
\end{equation}
The backbone model can also be rewritten as 
\begin{equation}
\hat{\mathbf{y}} = \mathbf{\Phi}_{\phi}(\eta(\mathbf{E}^{-24\mathrm{h}}), \mathbf{x}, \mathbf{c}),
\label{eq:ranker_formulation}
\end{equation}
where $\eta(\cdot)$ is the user-tower encoder that only has DV365 user timelines as input, and $\phi$ represents the rest of the parameters in $\mathbf{\Phi}$ other than $\eta$. The parameters in $\eta$ and $\phi$ are continuously trained in an recurring fashion, and the resulting user embedding is produced by evaluating the user tower $\eta(\mathbf{E}^{-24\mathrm{h}})$ across all active users on a 6-hourly basis. 

\subsubsection{Funnel Summarization Arch (FSA)}\label{sec:user_tower}

Design goal of DV365 user encoder $\eta(\mathbf{E})$ is to encode the multi-sliced embeddings into a more compact user embeddings as a representation learner plus compressor for lower infra and downstream consumption cost.

First, we apply a token-wise view at raw embeddings. Given a user embedding tensor of size $[N, D]$, where $N$ is the number of raw embedding after sparse lookup, and $D$ is the embedding dimension. We transpose the tensor to be $\mathbf{U}$ of size $[D, N]$ and apply neural networks on the last dimension (token dimension). Compared to the traditional way of viewing it as $[1, N\times D]$ in MLP layers or $[N, D]$ in transformers, this transpose enforces parameter sharing across all dimensions of the raw (pooled) token embeddings and thus preserves the semantic meaning of embedding in the output. 

Second, we apply Funnel Transformer \cite{Dai2020FunnelTransformerFO} encoder on the token-wise view. 
The basic transformer encoder layer can be expressed as  
\begin{equation}
\begin{split}
& \text{EncoderLayer}(\mathbf{U}) := \\
& \ \ \ \ \ \text{Norm}(\mathbf{U} + \text{FF}(\text{Norm}(\mathbf{U} + \text{Att}(q,k,v=\mathbf{U}))))
\end{split}
\label{eq:funnel_transformer_block}
\end{equation}
where FF is the position-wise feedforward network, Att is multihead self attention, Norm is layer-normalization. The funnel transformer can have multiple blocks with multiple layers in each block. 
Between each block, mean pooling is applied on the token dimension with a stride, so that $\mathbf{U}_{\text{out}}$ is of size $[D, N_{\text{out}}]$ where $N_{\text{out}} < N$. Importantly, the pooling is applied to the token dimension instead of the embedding dimension as in original Funnel Transformer \cite{Dai2020FunnelTransformerFO}. 
A Funnel Transformer encoder with $n$ blocks, each block having $m$ layers is described in Algorithm \ref{alg:FSA}. 
In our experiments, the funnel transformer achieves consistent NE performance against the regular transformer when the regular transformer is used as the token-wise setup, but it achieves a higher training speed due to fewer parameters (section \ref{sec:summarization_arch_experiments}). 

\begin{algorithm}
\caption{Funnel Transformer Encoder}\label{alg:FSA}
\begin{algorithmic}
\Require $\text{stride} \in \mathbb{Z}^+$
\Require $n, m \in \mathbb{Z}^+$ \Comment{Number of blocks and attention layers per block}
\Require $\text{EncoderLayer}_{i, j}, 1 \le i \le n, 1 \le j \le m$
\For{$i \gets 1$ to $n$}
\If{$i == 1$} \Comment{Don't pool before the 1st block}
    \State $\mathbf{U}' \gets \mathbf{U}$
\Else
    \State $\mathbf{U}' \gets \text{Pool}(\mathbf{U}, \text{stride})$  \Comment{Pool in token dimension}
\EndIf
\For{$j \gets 1$ to $m$}
\State $\mathbf{U}' = \text{EncoderLayer}_{i, j}(\mathbf{U}')$
\EndFor
\EndFor
\State \Return $\mathbf{U}'$
\end{algorithmic}
\end{algorithm}

We also add a linear compression encoder (LCE) in parallel to the funnel transformer for better performance. During training, output embeddings are stacked and flattened as part of the simulation network's inputs.
During publishing, they are quantized with 4-bit quantization before entering the feature store for serving. In total, the DV365 user tower (FSA and 4-bit quantization) compresses $200 \times 256$ fp32 numbers into $58 \times 17$ long integers, achieving a compression factor of 50. 

\subsection{Downstream Integration}

To integrate with downstrewam models, we need to de-quantize back to float embeddings and do a learnable projection to make DV365 embeddings (58 of 256 dim embeddings). What we adopted can be summarized in 3 types:

{\begin{enumerate}
    \item \textbf{DLRM ranking models:} After a linear projection, we concatenate DV365 embeddings with other sparse embeddings originally in the model.
    \item \textbf{HSTU module:} We prepend DV365 embeddings in the beginning of original HSTU input sequence (latest user engagements)
    \item \textbf{Retrieval (Siamese network or MoL\cite{Zhai2023RevisitingNeutralRetrieval}):} We empirically found GateNet \cite{Huang2020GateNetGD} is more effective than linear projection in retrieval integration.
\end{enumerate}}

\subsection{Serving in Production}

\begin{figure*}[ht] 
    \centering 
    \includegraphics[width=\textwidth]{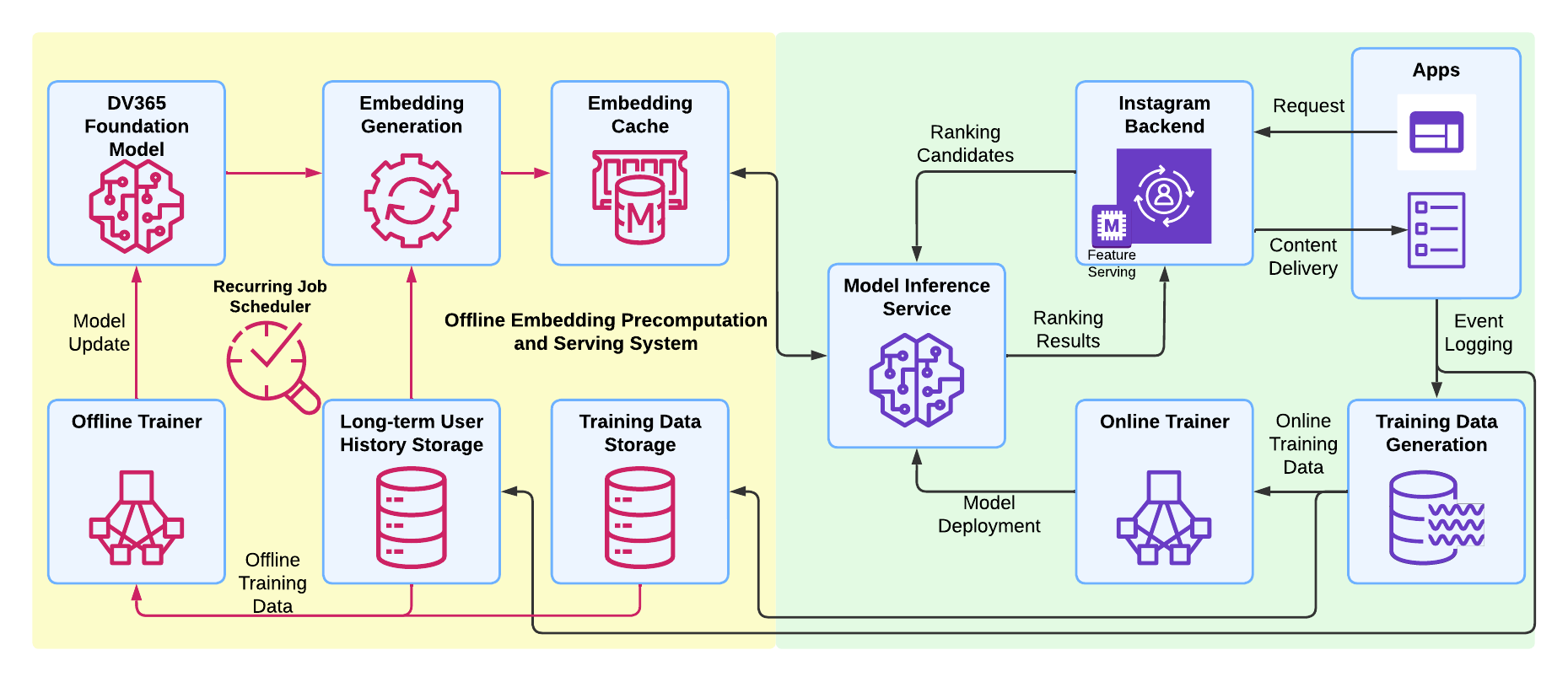} 
    \caption[Embedding Precomputation and Serving System]{
        Architecture of the recommendation system architecture incorporating DV365 embeddings. New components introduced are indicated in light yellow. 
    } 
    \Description[Embedding Precomputation and Serving System]{ 
        Diagram depicting the embedding precomputation and serving system, showcasing components for data collection, training, embedding generation, and caching. 
    } 
    \label{fig:dv365_e2e_system} 
\end{figure*}

\subsubsection{Existing System} 
The existing recommendation system initiates when a user sends a request through the Instagram frontend application. Upon receiving this request, the backend system prepares recommendation candidates and their associated features. These candidates are forwarded to the model inference service, where they are ranked based on predicted relevance. The ranked content is returned to the backend and delivered to the user. User interactions with the recommended content generate interaction events, which are logged to create training data for the core ranking model deployed in the inference service. The core ranking model employs an online training \cite{Online_Machine_Learning_2024} approach, continuously updating model weights with new data to maintain freshness with minimal delay. The components of this system are depicted in Figure \ref{fig:dv365_e2e_system}, highlighted in light green.

\subsubsection{Embedding Precomputation and Serving System} 
Long-term user interests and behavioral patterns remain relatively stable over short timeframes, such as several hours or days. As a result, frequent computation of user embeddings from raw interaction data is unnecessary and computationally expensive. So we propose an offline embedding precomputation and serving system, as illustrated in Figure \ref{fig:dv365_e2e_system}. The system incorporates the following key components and processes:

\begin{enumerate} 
    \item \textbf{Offline training data generation}: A data pipeline aggregates long-term user interaction histories and combines them with offline training data, forming a comprehensive dataset for training the DV365 foundation model.
    \item \textbf{Foundation model training}: The DV365 foundation model is trained on a recurring schedule using the aggregated data. This periodic training ensures that the model catches up with data distribution updates, despite being constantly behind the data consumed by the production online training by about 2 days. 
    \item \textbf{Embedding precomputation and serving}: User embeddings are generated by exporting intermediate outputs from the user tower of the trained model. These embeddings are indexed by user IDs, stored in a database, and cached in an online key-value store for efficient retrieval. Storage optimization techniques, such as embedding precision reduction and quantization, are employed to minimize resource usage while preserving embedding quality.
    \item \textbf{Recurring job scheduler}: A recurring job scheduler orchestrates the entire workflow, including data collection, model training, embedding generation, and caching. By operating at regular intervals, the scheduler ensures timely updates and balances computational efficiency with embedding freshness.
\end{enumerate}

The offline embedding system is a cost-efficiency highlight since it has lower requirement on keeping things real-time. We are able to update model less frequently, and long user history data are kept in low-cost data warehouse without being exposed to expensive online serving storage. Power consumption is further optimized by refreshing embeddings based on user activity levels and applying incremental updates to capture changes in user behavior.

\section{Experiments}

We perform extensive experiments to test the performance of DV365 embedding in the upstream ranker model with DV365 user tower, as well as downstream ranker and retrieval models. For ranker models, we use the normalized entropy (NE) as an offline evaluation metric for the ranker model's performance on binary tasks
\begin{equation}
\text{NE}_k = \frac{\text{BCEloss}(\hat{y}^{(k)}, y^{(k)})}{-\left[p_k\ln(p_k) + (1-p_k)\ln(1-p_k)\right]}, 
\label{eq:normalized_entropy}
\end{equation}
where $p_k$ is the average CTR (click-through rate) for task $k$. The normalization in the denominator helps to prevent the metric from being artificially good when the background CTR is close to 0 or 1 \cite{He2014NormalizedEntropy}. Instead of absolute NE score, we use \textbf{Relative NE Delta} computed on exact same date range for test and control because data distribution and background CTR change over time. At Meta, this metric is widely regarded as one of the most reliable offline metrics, whose movement is consistent with online A/B test results across recommendation and Ads ranking. \textbf{>0.1\% is considered significant and indicate A/B test gains.}

For retrieval models, we report hit rate @1 and @10 as evaluation metrics with a multitask version of MoL \cite{Zhai2023RevisitingNeutralRetrieval} but with simple dot-product as the user-item similarity function.

\subsection{Dataset \& Experiment Setup}
Our dataset is Instagram's internal training data produced from recommendation logs. In existing production training table (used to train existing production models, or referred as downstream models for DV365), each training row consists of user engagements as labels and features of user history and items. User history length in the production training table is on average around 1500 and 2000 at maximum.

In order to train DV365 on extremely long user sequence, we created another table which append longer history on top of production training table offline which is more cost-efficient than appending directly in real-time generated production table. User history in this table is on average around 40,000 and 70,000 at maximum.

The DV365 foundation model described in section 2 is trained on the offline long history table, and we make our model decisions based on NE metrics on the foundation model because we assume improvement in foundation model will transfer to downstream models through exported user embeddings.

Downstream models for DV365, or real production models, are diverse in objectives (retrieval or ranking) and verticals (Reels, Feed, Notification). In this section, we report the downstream performance trained on production table in our biggest vertical, which is Reels, on both ranking and retrieval. The ranking model baseline is a DLRM model and retrieval model baseline is a MoL\cite{Zhai2023RevisitingNeutralRetrieval}) model. Both of them use HSTU as their user sequence encoder, and were the production model at the time. This established a strong baseline because they are real production model with an already long user sequence ($O(10^3)$) and an attentive encoder HSTU. We measure additional gains when DV365 integrated as features in these baselines.

\subsection{Choice of Sequence Encoder}
\label{subsec:sequence_model_choice}

We choose mean-pooling as our sequence aggregation method given the scalability challenge of other methods in dealing with extremely long sequence. We experimented with attentive sequence models but didn't find substantial gain on top of our Multi-slicing \& Summarization baseline, details discussed in appendix \ref{appendix:sequence_model_choice}.

\subsection{Choice of Backbone Simulation Network}

We hypothesize consistency of upstream and downstream models are important for the knowledge transfer. So we tested 2 model backbone choices: DLRM, consistent with all of our production ranking models; and siamese network, consistent with our retrieval models. DLRM backed upstream showed similar performance in downstream retrieval models, but performed significantly better in downstream ranking models (0.2\% NE delta). So we choose DLRM as our backbone.

We also tested other factors in the simulation network:
\begin{enumerate}  
    \item \textbf{Tasks}: Unimportant. Including only a subset of critical downstream prediction tasks in the backbone maximizes the DV365 embedding’s effectiveness, while adding non-critical tasks yields minimal additional NE gains. 
    \item \textbf{Model Scales}: Unimportant. Increasing the complexity of task-specific modules or interaction architectures provides limited additional benefits to the DV365 embedding's downstream performance. 
    \item \textbf{Features}: Both sparse and dense features are important for simulation (0.2\%+ NE difference with complete removal). But as expected, embedding quality win from more simulation features diminishes after enough features.
\end{enumerate}

\subsection{Choice of DV365 User Encoder Architecture }\label{sec:summarization_arch_experiments}

\begin{table}[htb]
  \caption{FSA experiment setup. $\text{N}_\text{out}$ means output embedding number }
  \label{tab:FSA_experiment_setup}
  \begin{tabular}{lll}
    \toprule
    Name            & User tower setup                  & $\text{N}_\text{out}$ \\
    \midrule
    baseline        & no DV365 timelines                & 0 \\
    pass\_thru      & no compression                    & 200 \\
    LCE             & Linear compressed embedding       & 58 \\
    $\text{MLP}_d$  & Dim-wise MLP                      & 58 \\
    $\text{MLP}_t$  & Token-wise MLP                    & 58 \\
    $\text{T-L}_d$  & Dim-wise Transformer + LCE        & 58 \\
    $\text{T-L}_t$  & Token-wise Transformer + LCE      & 58 \\
    $\text{FSA}$    & Funnel Transformer + LCE          & 58 \\
  \bottomrule
\end{tabular}
\end{table}

\begin{table*}
  \caption{FSA compression performance compared to baseline in NE and MSE percentage delta (\%). More negative means better performance. The best 2 NE among the experiments are in \textbf{bold}.}
  \label{tab:FSA_experiment_results}
  \begin{tabular}{l|lllllll}
    \toprule
    Task     & pass\_thru & LCE & $\text{MLP}_d$ & $\text{MLP}_t$ & $\text{T-L}_d$ & $\text{T-L}_t$ & $\text{FSA}$ \\
    \midrule
    video\_complete     &-0.593          &-0.609 &-0.486 &-0.641 &-0.500 &\textbf{-0.659} &\textbf{-0.650} \\
    skip                &-0.480          &-0.497 &-0.395 &-0.495 &-0.398 &\textbf{-0.519} &\textbf{-0.525} \\
    reshare             &-0.348          &-0.327 &-0.185 & 0.056 &-0.169 &\textbf{-0.342} &\textbf{-0.355}  \\
    profile\_visit      &\textbf{-0.507} &-0.412 &-0.270 &-0.433 & 0.394 &-0.475          &\textbf{-0.480}  \\
    external\_share     &\textbf{-3.351} &-2.811 &-1.420 &-2.420 &-2.751 &\textbf{-3.273} &-3.159           \\
    follow              &\textbf{-1.034} &-0.801 &-0.421 &-0.858 &-0.697 &\textbf{-0.966} &-0.945           \\
    liked               &-0.573          &-0.564 &-0.268 &-0.580 &-0.191 &\textbf{-0.637} &\textbf{-0.634}  \\
    save                &\textbf{-0.795} &-0.648 &-0.206 &-0.702 &-0.005 &\textbf{-0.739} &-0.729           \\
    comment             &\textbf{-2.132} &-1.617 &-1.025 &-1.600 &-1.272 &\textbf{-2.004} &-1.927           \\
    $\text{watch\_time}_\text{MSE}$        &-0.830          &-0.848 &-0.632 &-0.843 &-0.664 &\textbf{-0.880} &\textbf{-0.865}  \\
  \bottomrule
\end{tabular}
\end{table*}

The goal of user encoder is feature interaction learning and compressing multi-sliced embeddings into deployable sizes. There are 2 design choices in the user encoder:
\begin{enumerate}
    \item \textbf{Compression perspective:} Dim-wise (apply encoder on [1, N x D] or [N, D]) or token-wise (transpose to [D, N] then encode)
    \item \textbf{Encoder architecture:} Linear Compression Encoder (LCE), MLP, Vanilla Transformer (take last $\text{N}_\text{out}$ output embeddings), Funnel Transformer \cite{Dai2020FunnelTransformerFO}
\end{enumerate}

We also add a no compression baseline in the experiments. The experiment setup is in table \ref{tab:FSA_experiment_setup}, and results are in table \ref{tab:FSA_experiment_results}. Results shows token-wise compression significantly outperforms dim-wise in all encoders, which indicate model favors parameter sharing within a single token embedding. Regarding to encoder choice, vanilla Transformer encoder and FSA yields best performance among compressed candidates. We chose FSA because it has fewer parameters and has 10\% training QPS advantage.

\subsection{Downstream Integration \& Product Launches}

For reporting convenience, we report offline result of only Reels surface (biggest engagement traffic), other surfaces yields similar results. For ranking model, DV365 improves NE significantly across all tasks (average over 0.4\%) (Table \ref{tab:ranking_experiment_result}) and linear projection results in better adoption. For retrieval model, DV365 increases hit rate by 2\% to 8\%, with GateNet as adoption module yields more gain in most tasks (table \ref{tab:retrieval_experiment_results}). 

DV365 has been widely launched in \textbf{15} different product models across Instagram and Threads with significant business impacts. Accumulating A/B testing from all launches, it has improved \textbf{0.7\% of Instagram App time spent} along with many other critical engagement metrics. It's highly generalizable: Despite highly heterogeneous downstream model architectures and objectives, we are able to use only one upstream model that generate a single set of embeddings to serve diverse downstream use cases, including improving content recommendation quality, notification CTR and jump starting new business initiative such as Threads.

We also verified that our design overcame the freshness disadvantage of DV365 as an offline embedding, i.e., embedding quality remains stable as staleness increases (details in Appendix \ref{appendix:robustness_to_staleness}). This is also a system reliability highlight considering DV365 is a critical dependency of multiple products in Instagram.

\begin{table}[htb]
  \caption{Downstream Ranking downstream NE difference in percentage delta (\%). More negative the better.}
  \label{tab:ranking_experiment_result}
  \begin{tabular}{l|lll}
    \toprule
    Task                & no projection         & linear projection          \\
    \midrule
    video\_complete     & -0.564       & \textbf{-0.637} \\
    skip                & -0.447       & \textbf{-0.508} \\
    reshare             & -0.077       & \textbf{-0.179} \\
    profile\_visit      & -0.261       & \textbf{-0.394} \\
    follow              & -0.282       & \textbf{-0.592} \\
    liked               & -0.455       & \textbf{-0.590} \\
    save                & -0.176       & \textbf{-0.356} \\
    comment             & -0.574       & \textbf{-1.151} \\
    $\text{watch\_time}_\text{MSE}$         & -0.737       & \textbf{-0.835} \\
  \bottomrule
\end{tabular}
\end{table}

\begin{table}[htb]
  \caption{Retrieval downstream adoption hit rates (HR). More positive means better performance.}
  \label{tab:retrieval_experiment_results}
  \begin{tabular}{l|lll}
    \toprule
    Task HR                & base  & DV365     & $\text{DV365}_{\text{GateNet}}$ \\
    \midrule
    follow@1             & 0.413     & 0.431 (4.4\%) & \textbf{0.437 (5.8\%)} \\
    follow@10            & 0.792     & 0.796 (0.5\%) & \textbf{0.802 (1.3\%)} \\
    liked@1             & 0.354     & 0.370 (4.5\%) & \textbf{0.370 (4.5\%)} \\
    liked@10            & 0.810     & 0.827 (2.1\%) & \textbf{0.830 (2.5\%)} \\
    save@1              & 0.423     & 0.439 (3.8\%) & \textbf{0.446 (5.4\%)} \\
    save@10             & 0.852     & 0.863 (1.3\%) & \textbf{0.874 (2.6\%)} \\
    reshare@1           & 0.316     & \textbf{0.341 (7.9\%)} & 0.337 (6.6\%) \\
    reshare@10          & 0.788     & \textbf{0.815 (3.4\%)} & 0.810 (2.8\%) \\
    profile\_visit@1    & 0.374     & 0.398 (6.4\%) & \textbf{0.402 (7.5\%)} \\
    profile\_visit@10   & 0.811     & 0.830 (2.3\%) & \textbf{0.836 (3.1\%)} \\
    watch\_time@1       & 0.243     & 0.260 (7.0\%) & \textbf{0.261 (7.4\%)} \\
    watch\_time@10      & 0.710     & 0.735 (3.5\%) & \textbf{0.738 (3.9\%)} \\
  \bottomrule
\end{tabular}
\end{table}

\section{ROI Advantages of Offline Embedding}
We compare cost of achieving the same methodology used in DV365 as offline embedding and as part of E2E model. In Meta, we estimate cost by normalizing power usage in data center using Watt.

\subsection{Feature Infra Cost Saving}
Compared to online E2E long sequence modeling methods, key advantage of offline embedding is it saves significant cost from real-time feature extraction and processing for model inference, and cost from storing the extremely long user history in low latency key-value storage. We measured in details that delta of the two approaches is around 100 MW from feature extraction cost, and 11 PB  low latency key value storage (Appendix \ref{appendix:k-v-store-saving}). This extremely high cost is the reason why extremely long user history is currently inaccessible in Meta's real-time serving infra.

\subsection{Model Training Cost Saving}
Model training cost saving comes from a single foundational model doing computation for all downstream models (15 in our case). In IG Reels model, additional feature preprocessing from long-history timelines to sliced features is measured 50 KW and GPU cost additional 338 A100 pool reservation. Considering not every model cost the same, we multiply total cost factor as 10 instead of 15, which is 500KW and 3380 A100 GPU reservation.

\subsection{Model Serving Cost Saving}
During model serving, feature preprocessing to slice timeline into features (saved by offline embedding) cost 6 MW and additional inference GPU cost (from user encoder) is 78 H100 (Appendix \ref{appendix:serving-saving}) for IG Reels model, and estimated 780 H100 additional across fleet.

\section{Related Works}

\subsection{Sequence User Modeling}

User sequence modeling is a classic problem for recommendation, and industry state-of-the-art has converged to attentive methods where an attention weight was learnt for each item in the user sequence to aggregate full user interest representation. Notable examples are DIN \cite{zhou2018din}, DIEN \cite{zhou2018dien}, SASRec \cite{kang2018selfattentivesequentialrecommendation}, HSTU \cite{zhai2024actionsspeaklouderwords}. And HSTU is an important module currently deployed in Instagram production models. These attentive methods have a well known sequence length scalability challenge that most deployed solutions are only at O(100) level. 

\subsection{Long-term User Sequence Modeling}

MIMN \cite{Pi_2019_MIMN} and HPMN \cite{Ren_2019_HPMN} use memory networks that improve deployable history access length to O(1000) but find it difficult scaling beyond that. Another theme of works (SIM \cite{qi2020searchbasedusermodelinglifelong}, SDIM \cite{cao2022samplingneedmodelinglongterm}, TWIN \cite{chang2023twin}) introduces the two stage end-to-end cascaded framework to enable longer term history modeling. The two stages are: 1) a fast search unit which retrieves the most “relevant” items to the target item from thousands of user engagements, 2) an attentive network to perform target attention over the finalists from stage 1. These works claimed to support deployed solutions that model \(10^4\) user sequences. Despite the big improvement, these methods still significantly increased the cost and complexity of existing production models by introducing the search unit for both training and serving GPU, and more importantly, it requires serving infra to access the long user history feature in real-time, which is a big additional feature infra cost. Furthermore, as an end-to-end solution, their learnt long-term user interest can only benefit one model at a time without wide knowledge sharing.

\subsection{User Embedding}

Offline user embedding is another way to offload online computation to offline batch jobs for higher cost efficiency. Pinnerformer \cite{pancha2022pinnerformersequencemodelinguser} introduced a transformer based upstream model to precompute user long-term interest with an offline embedding which is successfully deployed in Pinterest with high business impact, however, since it uses expensive encoder and only report to deploy max sequence length 256. LURM \cite{yang2023empoweringgeneralpurposeuserrepresentation} introduced a bag-of-interest method to use pre-learned clustering to compress long history to a vector of per-cluster count, and used a self-supervised training objective to learn user embedding for generalization. Bag-of-interest is a very cost-efficient compression, but introduces a system dependency which is the clustering learning and the paper doesn't share A/B test and deployment results.

\section{Conclusion}

We proved our new long user history focused user embedding with high deployment success and strong ROI justifications. Specifically, our multi-slicing data model and user embedding summarization framework is able to capture sufficient information from the long user history and yield significant improvements in downstream production models equipped with SOTA model architectures. With this strategy, we are able to deploy the DV365 embedding to tens of downstream models with significant topline impact at Instagram. 

\newpage

\begin{acks}
We highly acknowledge Max Kaplan, Parichay Kapoor, Dhruv Choudhary, Hanxiong Chen, Zhengze Zhou, Mark Gluzman, Hanxiong Chen, Gufan Yin, Keyi Wu, Sami Bendouba, Xiaoyang Gong, Enzhou Liu, Ziheng Huang, Erheng Zhong, Joyce Liu for hands-on contributions, Ji Liu for insightful discussions, and Maxime Ransan, Arun Kumar Singh, Misael Manjarres, Nathan Berrebbi, Deepti Chheda, Abhishek Kumar, Haotian Wu for sponsorship and management support.
\end{acks}

\bibliographystyle{ACM-Reference-Format}
\bibliography{sample-base}

\appendix

\section{Detailed Cost Estimation}
\subsection{Stats of User History Length Modeled}
Implicit timelines (user impression history): Max and average length are 35K and 30K; 
Explicit timelines (user explicit action history): Max and average length are 35K and 10K. The 35K maximum is a parameter we chose based on ROI. So potential total length is 70K in maximum and 40K on average.

\subsection{Details of Feature Serving Infra Cost Estimation}
\subsubsection{Feature Extraction Cost} \label{appendix:feature_extraction_saving}
\begin{enumerate}
    \item \textbf{Serving long sequence online} For cheaper estimation, we did a online load testing of extracting average length 60 user history in our feature infra, measured as 150 KW power, and estimate full DV365 (40K average) cost by multiply length ratio, which is 150 KW * (40K / 60) = 100 MW.
    \item \textbf{Serving DV365 embedding:} 21 KW as measured by production load testing.
\end{enumerate}

\subsubsection{Key-value Storage}\label{appendix:k-v-store-saving}
\begin{enumerate}
    \item \textbf{Serving long-sequence online:}  8 bytes (int64 type) * ((30K average length * 12 implicit attributes) + (10K average length * 10 explicit attributes)) * 3B (users)  = 11 PB.
    \item \textbf{Serving DV365 embedding:} With int4 quantization, each output embedding takes 136 bytes and there are 58 output embeddings in total. Therefore, the total storage is 136 bytes / embedding * 58 embedding * 3B (users) = 24 TB.
\end{enumerate}

\subsection{Details of Model Training Cost Saving}\label{appendix:training-saving}
Model training cost is the same for E2E and offline model training if there is only one consumer model of the offline embedding. Our saving mainly comes from one offline model that serves 15 downstream models.

Introducing DV365 components in a single model's cost has 2 parts:
\begin{enumerate}
    \item \textbf{Feature Preprocessing from Long-history timeline to sliced features:} 1K CPU hosts * 3 hour (training time) * 100 W / CPU host * 4 (training occurrence per day) / 24 hour (normalized over a day) = 50 KW.
    \item \textbf{Training GPU:} Additional model components in DV365 cost training QPS regression that translate to 338 A100 cards in our resource pool for both production and experimental training added together. 
\end{enumerate}

\subsection{Details of Model Serving Cost Saving} \label{appendix:serving-saving}

Additional cost of E2E method modeling long sequence mainly comes from feature pre-processing computed in CPU hosts before sending processed features to GPU inference. We did an online A/B testing and measured 900KW to run DV365 feature preprocessing in user history average length of 60. Since we assume there is big optimization space, we apply a generous 0.01 optimization factor for fair estimation. So final estimation is 900KW * (40000 length we use / 60 tested length) * 0.01 optimization space factor = 6MW

Additional GPU inference cost comes from Funnel Summarization Arch, which cost inference QPS regression that translates to 78 H100 cards in IG Reels. 

\section{Robustness to staleness}\label{appendix:robustness_to_staleness}

DV365 embeddings are designed to capture users' long-term, stable interests, making them inherently resilient to delays and staleness. This robustness ensures their effectiveness under variable update cadences, a crucial requirement for real-world applications.

To systematically assess the impact of staleness on model performance, we conducted a controlled experiment involving three downstream models: (1) a baseline model without DV365 embeddings, (2) a model with daily updated DV365 embeddings (referred to as "fresh embeddings"), and (3) a model using a fixed, non-updating version of the embeddings (referred to as "stale embeddings"). All models were trained on seven days of data, and the NE gain relative to the baseline model was continuously monitored.

\begin{figure}[ht] 
    \centering 
    \includegraphics[width=0.47\textwidth]{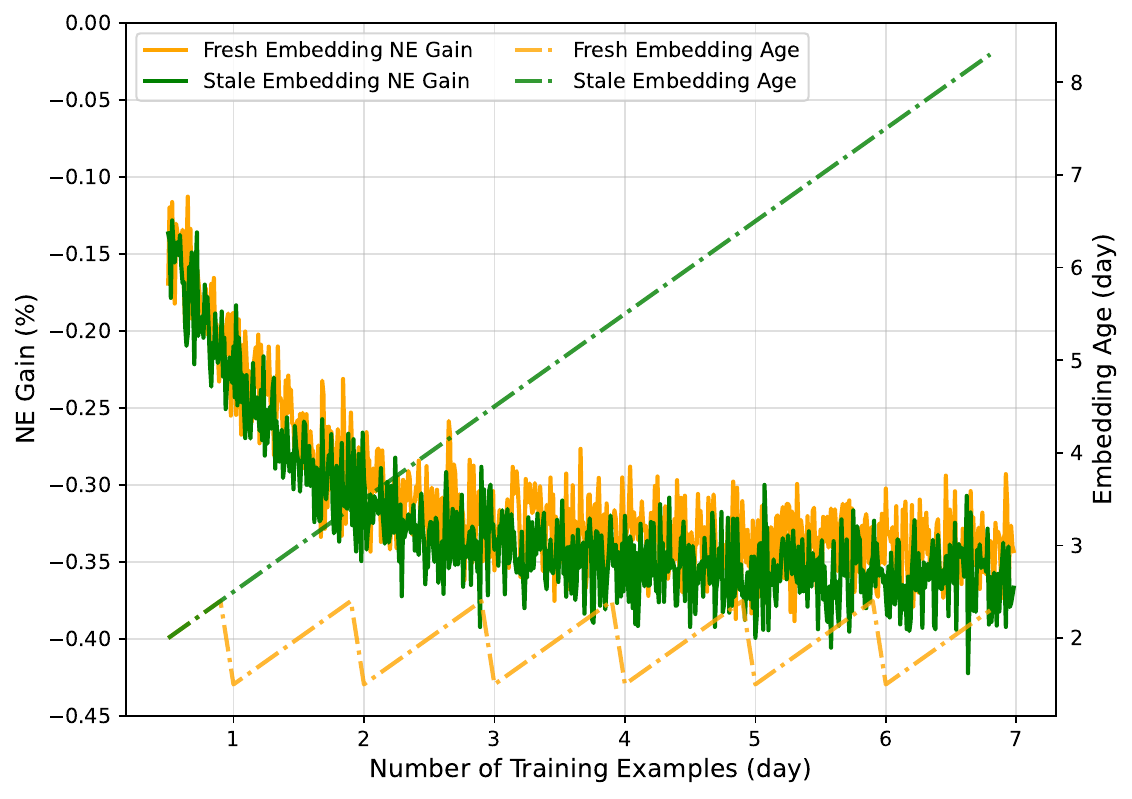} 
    \caption[Embedding Robustness to Staleness]{
        This figure illustrates the relationship between embedding NE gain (relative to the baseline model) and the number of training examples. The x-axis represents the cumulative number of training examples, measured in days. For the regularly updated embeddings, staleness fluctuates periodically due to daily updates. In contrast, the staleness of the fixed embeddings increases steadily over time, as these embeddings are not updated. The NE gain initially increases over the first three days of training data and then converges, stabilizing at a consistent level.
    } 
    \Description[Embedding Robustness to Staleness]{Embedding Robustness to Staleness} 
    \label{fig:dv365_embedding_robustness} 
\end{figure}

Throughout the training period, the staleness of the fresh embeddings remained relatively stable, averaging around two days due to the daily updates. In contrast, the staleness of the stale embeddings increased steadily over time. Despite this disparity, both the fresh and stale embedding models demonstrated consistent NE gains of approximately 0.35\% compared to the baseline, as shown in Figure \ref{fig:dv365_embedding_robustness}. These results underscore the DV365 embeddings' robustness to staleness, ensuring reliable performance even when updates are delayed or entirely absent.

\begin{figure}[ht] 
    \centering \
    \includegraphics[width=0.47\textwidth]{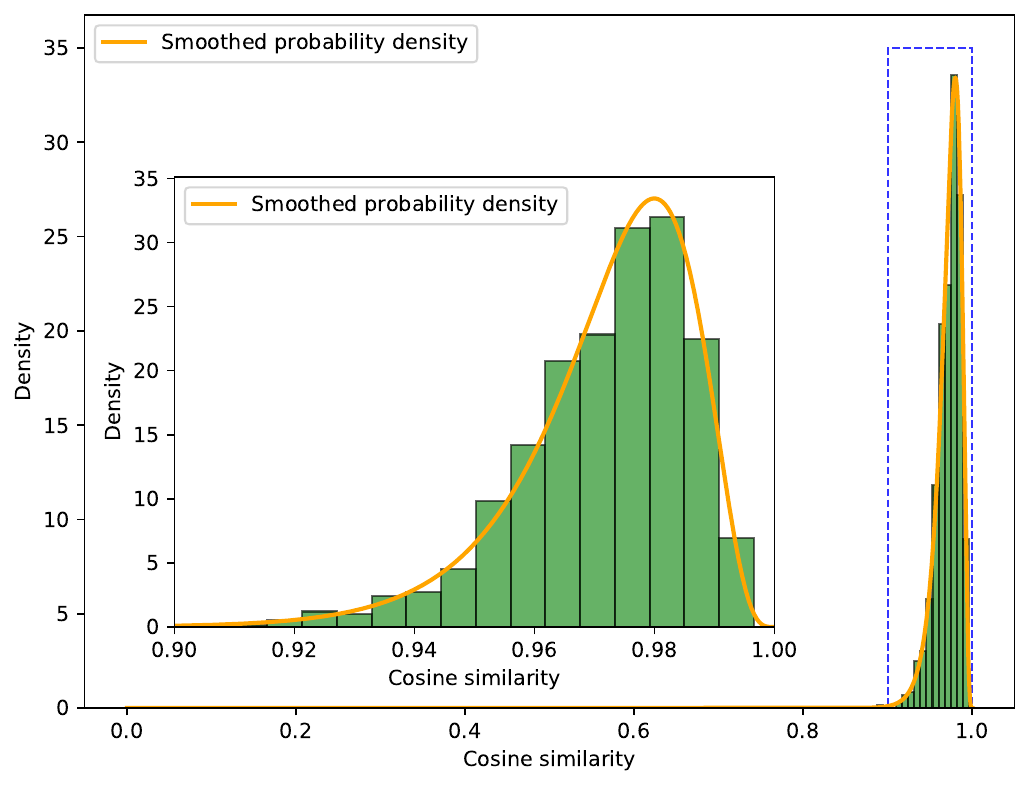} 
    \caption[User Embedding Similarity]{ 
        Distribution of cosine similarity between two user embedding versions. The similarity distribution is sharply unimodal, centered close to 1. The inset chart provides a closer view of the cosine similarity range between [0.9, 1.0]. 
    } 
    \Description[User Embedding Similarity]{User Embedding Similarity} 
    \label{fig:dv365_user_emb_similarity} 
\end{figure}

To further assess the stability of the DV365 embeddings over time, we analyzed the cosine similarity between different embedding versions. Remarkably, even when two versions were separated by seven days, the cosine similarity consistently exceeded 90\%. This indicates that the embeddings maintain a high degree of stability over extended periods, as shown in Figure \ref{fig:dv365_user_emb_similarity}.

These findings highlight that the DV365 embeddings not only resist performance degradation caused by staleness but also exhibit exceptional consistency in representing user interests. This stability and reliability stem from the design of the DV365 embeddings, which leverage extensive user history to capture long-term patterns effectively, ensuring dependable performance across downstream models.

\section{Choice of Sequence Encoder}
\label{appendix:sequence_model_choice}

We choose mean-pooling as our sequence aggregation method given the scalability challenge of other methods in dealing with extremely long sequence. But we still conduct experiments on attentive sequence models to understand its headroom on top of our multi-slicing pooling baseline:

\begin{enumerate}
    \item \textbf{Position-weighted pooling:} As a sequence model baseline, we simply learn a positional weight for each item in user sequence, and use it to apply weighted pooling.
    \item \textbf{HSTU \cite{zhai2024actionsspeaklouderwords}:} We added a HSTU encoder that encode user sequence into a summarized embedding with self attention. Given semantic similarity between HSTU and Transformer, we believe conclusion of this implementation can also represent Transformer based sequence encoders.
\end{enumerate}

Our experiment results show both options is quite incremental if overall user sequence length is relatively (we apply a sequence length cap in the baseline to simulate this situation), but as we scale sequence length to 35000 for both explicit and implicit timeline in the baseline multi-slicing implementation, gains coming from additional sequence model encoded embedding fully diminished. (For HSTU, due to its scalability limitation we scaled it to 1000 instead of 35000).

Based on this observation, we decide to drop attentive sequence pooling in DV365 upstream model at the moment. We hypothesize that sequential information is mainly useful for user's emerging interest, but not so much on long-term stable interest. For stable interest, counting based summarization (captured in mean pooling) represents a strong baseline.

As mentioned in \ref{eq:time_slicing}, we apply a cost-neutral time-slicing on the sequence before apply mean pooling, which gives us 0.1\% NE gain in many tasks.

\end{document}